\documentstyle[12pt]{article}
\newcommand{\beq}{\begin{equation}}
\newcommand{\eeq}{\end{equation}}
\newcommand{\bea}{\begin{eqnarray}}
\newcommand{\eea}{\end{eqnarray}}

\newcommand{\goto}{\rightarrow}
\newcommand{\be}{\beta}
\newcommand{\zb}{\bar{z}}

\newcommand{\al}{\alpha}

\begin{document}
\topmargin 0pt
\oddsidemargin 5mm
\renewcommand{\thefootnote}{\fnsymbol{footnote}}
\newpage
\setcounter{page}{0}
\begin{titlepage}\begin{flushright}
OU-TP-97-25P\\
{\bf hep-th/9705240}\\
 {\it May 1997}
\end{flushright}
\vspace{0.5cm}
\begin{center}
{\Large {\bf Origin of Logarithmic Operators in 
Conformal Field Theories}} 
\\
\vspace{1.8cm}
\vspace{0.5cm}
{Ian I. Kogan\footnote{e-mail:
i.kogan1@physics.oxford.ac.uk} and  Alex
Lewis\footnote{e-mail:
a.lewis1@physics.oxford.ac.uk} \\}
\vspace{0.5cm}
{\em Department of Physics, University of Oxford\\
1 Keble Road, Oxford, OX1 3NP, United Kingdom}\\
\vspace{0.5cm}
\renewcommand{\thefootnote}{\arabic{footnote}}
\setcounter{footnote}{0}
\begin{abstract}
{We study logarithmic operators in Coulomb gas models, and
show that they occur when the ``puncture'' operator of 
the Liouville theory is included in the model. We also consider
WZNW models for $SL(2,{\cal R})$, and for $SU(2)$ at level $0$, in
which we find 
logarithmic operators which form Jordan blocks for the current as
well as the Virasoro algebra.}
\end{abstract} 
\vspace{0.5cm}
 \end{center}
\end{titlepage}
\newpage
\section{Introduction}

 Logarithmic operators in conformal field theory were first studied
by Gurarie in the $c=-2$ model \cite{gur}.
   These logarithms 
 have now been found  in a multitude of other models such as the
WZNW model on the supergroup GL(1,1) \cite{rs}, the
gravitationally  dressed CFTs \cite{bilal}, 
$c_{p,1}$ and non-minimal $c_{p,q}$ models \cite{{gur},{f},{k},{gk}}, 
critical
disordered models \cite{{tsvelik},{ms}}, and WZNW models at level $0$
\cite{kogmav,cklt}, and  play a role in
the study of critical polymers and percolation \cite{{f},{k},{s},{w}},
 $2D$-magneto-hydrodynamic turbulence and ordinary turbulence 
\cite{rr,flohr} and  quantum Hall states \cite{flohr2,gfn}.
  They are also important for studying the problem of recoil
in the theory of strings  and $D$-branes
\cite{{kogmav},{kmw},{p},{emn},{texas}} as well as target-space symmetries
in string theory in general \cite{kogmav}. The representation theory
of
the Virasoro algebra for logarithmic CFT was developed in \cite{rohsiepe}.

In this paper we discuss the
free
field formulation of CFT with logarithmic operators, and we show that
they are closely related to the ``puncture'' operator of 2D gravity.
In sections 2, 3 and 4, we discuss $c_{p,q}$ models, and 
gravitationally dressed CFT. 
 In section 5, we consider the analogous
situation
in models with an affine Lie algebra 
as well as a Virasoro algebra. It was first
realised in \cite{gur} that when there are logarithmic operators, the 
generators of the Virasoro algebra cannot be diagonalized, but have a
Jordan cell structure:
\bea
L_0 |C\rangle &=& h |C\rangle \nonumber \\
L_0 |D\rangle &=& h |D \rangle + |C\rangle
\label{jordan}\eea
There are
two possible  types of Jordan cell structures that we can consider
in the analogous case when there is a 
current\footnote{The affine Lie algebra (current algebra) 
is often referred to as the Kac-Moody
algebra, althogh this may be misleading as the affine algebras were
discovered independently of the Kac-Moody algebra, and in the correct
mathematical nomenclature ``Kac-Moody algebra'' refers to a more
general case. For a short history of the subject see the appendix of
\cite{halpern}.}
 as well as a Virasoro
algebra. The first possibility, 
which is directly 
analogous to eq. (\ref{jordan}), is to have operators with the
behaviour (first introduced in \cite{oleg}):
\bea
J^a_0 C &=& t^a C \nonumber \\ 
J^a_0 D &=& t^a D + C \nonumber \\
J^a_n C = J^a_n D &=& 0,~~~~~~~~~~n\geq 1.
\label{KMjordan1}\eea
Perhaps surprisingly, the operators $C$ and $D$ with this behaviour
do not also form Jordan cells for the Virasoro algebra of the 
WZNW model (although they might in other models), so although
they have the same conformal dimensions, they are both primary, not 
logarithmic, operators of the Virasoro algebra. The second possibility
is to have a pair of operators which form a Jordan cell for $L_0$ but
not for $J^a_0$:
\bea
J^a_0 C = t^a C,~~~~~~~J^a_n C=0,~~~~~~~n\geq 1 \nonumber \\
J^a_0 D = t^a D,~~~~~~~J^a_1 D=t^a K,~~~~~~~J^a_n D =0,
~~~~~~~n\geq 2 \nonumber \\
J^a_0 K = t^a K,~~~~~~~J^a_n K=0,~~~~~~~n\geq 1.
\label{KMjordan2}\eea
In this case, $C$ and $D$ are a logarithmic pair,
 obeying eq. (\ref{jordan}), while $K$ is a primary field with
a dimension $1$ lower than that of $C$ and $D$. It is $D$ and $K$ 
that form a Jordan cell for $J^a_1$. Similar constructions
are possible where the dimensions differ by integers greater than $1$.

Let us begin by  reviewing
 how  logarithmic operators appear in conformal field theory.
 Let us consider for example
  the four-point correlation functions of a
 primary field $\mu(z)$  with anomalous dimensions $h$,
 such that $<\mu(z) \mu(0)> = z^{-2h}$. This correlation function
 can be represented as
\bea
<\mu(z_1) \mu(z_2) \mu(z_3) \mu(z_4)> =
\frac{1}{|(z_1 - z_2)(z_3 - z_4)|^{4h}}  
\sum_{i,j}U_{ij}F^i(x)F^j(\bar{x}),
\label{F}
\eea
where $x = (z_1 - z_2)(z_3 - z_4)/(z_1 - z_4)(z_3 - z_2)$.
 In many different models the
 unknown functions $F^i(x)$ (the conformal blocks)
are given  as  solutions of the
 hypergeometric equation 
(or a higher order Fuchsian differential equation).
\bea
x(1-x)\frac{d^2 {\cal F}}{dx^2} +
[d - (a+b+1)x] \frac{d{\cal F}}{dx} - ab {\cal F} = 0
\label{hypereq}
\eea
 which in general has two  independent solutions 
 \bea
{\cal F}_1 = F(a,b,d; x), ~~~~~~~
{\cal F}_2 = x^{1-d} F(a-d+1,b-d+1,2-d; x)
 \eea
where $F(a,b,d; x)$ is a hypergeometric function (we use $d$ as
the third parameter to avoid confusion with the central charge $c$)
and these two
independent solutions correspond to the  two primary fields in the
 OPE of
\bea
 \mu(z) \mu(0) = \frac{1}{z^{2h}}~[ I + z^{d-1} O  + \cdots]
\label{OPE1}
\eea
one of which is an identity operator $I$
 and the second  operator $O$   has an
 anomalous dimension $1-d$. It is simplest to discuss the case
when there are only two primary fields in the OPE, although in general
there are more; for example, in  minimal models,
the fields appearing in the OPE are given by
\beq
\phi_{r_1,s_1} \times \phi_{r_2,s_2} =
\sum_{r_3=|r_1-r_2|+1}^{r_1+r_2-1}
\sum_{s_3=|s_1-s_2|+1}^{s_1+s_2-1}
\phi_{r_3,s_3}
\label{minope}\eeq
 If $\mu$ in eqs. (\ref{F}) and (\ref{OPE1}) is the $(1,2)$
field in a minimal model, $O$ is the $(1,3)$ field and 
the conformal blocks are given by
eq. (\ref{hypereq}) with 
\bea
a &=& \frac{1}{24}\left[\sqrt{1-c} - \sqrt{25-c}
\right]^2\nonumber \\
b &=& -1 + \frac{1}{8}\left[\sqrt{1-c} - \sqrt{25-c}
\right]^2\nonumber \\
d &=& \frac{1}{12}\left[\sqrt{1-c} - \sqrt{25-c}
\right]^2\label{abc}\eea
In this way we recover the
 conformal blocks in a generic CFT \cite{bpz}.
 However, this is not true  if the parameter
  $d$ in the  hypergeometric equation (\ref{hypereq}) is an
 integer.  There  is a  general  theorem in  the theory
 of  the second order  differential equations which deals with
 the expansion of the solution near the regular point
   $x=0$
\bea
x^{\alpha}\sum_{n} a_n x^n
\label{ryad}
\eea
 This theorem  tells us how to calculate the coefficients $a_n$ if one
 knows  the two roots $\alpha_1$ and $\alpha_2$, which are the
 solutions of  the  so-called indicial equation \cite{ww}.
 However, if the difference $\alpha_1 - \alpha_2$
 is an integer, the  second solution either equals the first one
 (when $\alpha_1 =\alpha_2$) or some of the coefficients are undefined.
  In both cases the second solution  has
  logarithmic terms $x^n \ln x$ in the expansion (\ref{ryad}), besides
the usual terms $x^n$.  For the hypergeometric
 equation the indicial equation is
\bea
\alpha(\alpha - 1 + d) = 0
\eea
and the two roots are $\alpha_1 = 0$ and $\alpha_2 = d-1$, i.e.
 for integer $d = 1+ m$ the second solution has logarithmic terms.

 Note that the dimensions of primary fields that appear in
the OPE (\ref{OPE1}) are given by the roots of the indicial equation.
This can be seen by substituting the general OPE
\beq
\mu(z)\mu(w)= \frac{1}{(z-w)^{2h}}\sum\left\{(z-w)^{h_i}O_i(w)
+\cdots\right\},
\eeq
where $h_i$ is the dimension of $O_i$, into the four-point function,
giving (omitting the $\bar{z}$ dependence)
\beq
<\mu(z_1) \mu(z_2) \mu(z_3) \mu(z_4)> =
\frac{1}{(z_1 - z_2)^{2h}(z_3 - z_4)^{2h}}
\sum_i \left(\frac{(z_1-z_2)(z_3-z_4)}{(z_1-z_4)(z_2-z_3)}
\right)^{h_i}+\cdots.
\eeq
Comparing this with eq. (\ref{F}), we see that:
\beq
F_i(x) = x^h_i(1 +\dots)
\eeq
So that $h_i$ is given by $\al$ in eq. (\ref{ryad}), which is
a root of the indicial equation. 
 
Therefore the
logarithmic solutions occur when the dimension of the
second operator $O$ in the OPE
(\ref{OPE1}) is degenerate either with the identity operator
 (when $m=0$ and $d=1$) or  with  one of its Virasoro descendants
 (for negative integer $m$ when $O$ has a positive  dimension
 $|m|$). For positive  integer $m$ (in which case $O$ itself
 has a negative dimension $-m$) one of its descendants will be
 degenerate with $I$ (in both cases, the descendant in question
is a null vector of the Virasoro algebra, as is discussed in section 4). 
If we had considered a more general 
four-point function containing two distinct primary fields, 
we would have a similar situation with another primary operator
taking the place of the identity. If we had a higher order differential
equation
instead of the hypergeometric equation, the above discussion 
would apply whenever two of the roots of the indicial equation
differed by an integer; in all cases, logarithmic conformal blocks 
occur when the dimensions of two of the operators in an OPE 
become degenerate.

The condition that $d= 1+ m,~~m \in Z$ is necessary, but not
sufficient if $m \neq 0$. To have logarithms in this case
 one has to impose additional
 constraints on $a$ and $b$ \cite{ww},\cite{gr}.

 If $d= 1 + m$, where $m$ is a natural number,
  the two independent solutions  are
\bea
{\cal F}_1 = F(a,b,1+m; x), \nonumber \\
{\cal F}_2 = \log x~ F(a,b,1+m; x) + H(x),
\label{1+m}
 \eea
where $H(x) = x^{-m}\sum_{k=0}^{\infty}h_k x^k$ and $h_{m} =0$,
unless either $a$ or $b$ equal $1+m'$ with $m'$ a natural number
 $m'<m$. In this case  the second solution is only a polynomial in
$x^{-1}$ and the are no logarithms.

 If $d= 1 - m$, where $m$ is a natural number,
  the two independent solutions  are
\bea
{\cal F}_1 = x^m F(a+m,b+m,1+m; x), \nonumber \\
{\cal F}_2 = \log x~ x^m  F(a+m,b+m,1+m; x) + H(x),
\label{1-m}
 \eea
where $H(x)$ is again some regular expansion without logarithms,
unless either $a$ or $b$ equal $-m'$ with an integer $m'$
such that  $0 \leq m'<m$, in  which  case  ${\cal F}_1$ is
a polynomial in $x$ and there are no logarithms.

An example of a correlation function with logarithms
 is the four-point function of the $(1,2)$ operator
in the model with $c=-2$ \cite{gur}. In this case, eq. (\ref{abc})
gives $a=b=1/2$, $d=1$, and the conformal blocks are given by
eq. (\ref{1+m}). The full correlation function, satisfying the
constraints of crossing symmetry and locality, is:
\beq
<\mu(z_1) \mu(z_2) \mu(z_3) \mu(z_4)> =
|(z_1 - z_3)(z_2 - z_4)|^{-4h}|x(1-x)|^{-4h}
\left[F(x)F(1-\bar{x})  + F(1-x)F(\bar{x})\right]
\eeq
where $h=-\frac18$ and
\bea
F(x)&=&F\left(\frac12,\frac12,1;x\right) \nonumber \\
F(1-x)&=&\log x~F\left(\frac12,\frac12,1;x\right)+H(x)
\eea
As was shown in \cite{gur}, this leads to the OPE
\beq
\mu(z) \mu(0) = |z|^{1/2}\left[
D(0)+\log |z|^2 C(0) +\cdots\right] 
\label{OPE2}\eeq
$D(z)$ is the logarithmic operator. 
The behaviour of the logarithmic pair $C$ and $D$ is expressed
by the OPE's with the stress tensor:
\bea
T(z)C(0) &=& \frac{h}{z^2}C(0) + 
\frac{1}{z}\partial C(0) +\cdots \\
T(z)D(0) &=& \frac{h}{z^2}D(0) + 
\frac{1}{z^2}C(0) + \frac{1}{z}\partial D(0) 
\label{TD}\eea
This implies that, instead of the usual irreducible
representations, $C$ and $D$ and their descendants 
form an indecomposable representation of the Virasoro algebra,
with $C$ and $D$ forming the basis of a Jordan cell for $L_0$ 
(as in eq. (\ref{jordan})).

\section{Logarithmic and Pre-Logarithmic Operators in
Coulomb Gas Models}
We would like to understand the origin of logarithmic operators in models with
the Liouville action, which describes 2D gravity 
in conformal gauge as well Coulomb gas models with $c<1$:
\beq
S = \frac{1}{8\pi} \int d^2\xi \sqrt{g(\xi)}
\left[\partial_{\mu}\phi(\xi) \partial^{\mu}\phi(\xi) 
+ i\alpha_0R^{(2)}(\xi)\phi(\xi) \right].
\label{action} \eeq
The central charge of the above action is 
\beq
c=1-24 \alpha_0{}^2
\label{charge}\eeq
and the stress tensor is:
\beq
T(z)=-\frac{1}{4}:\partial_z\phi\partial_z\phi:+i\al_0\partial_z^2\phi
\label{T(z)}\eeq
The field $\phi$ has the short distance behaviour:
\beq
\phi(z)\phi(w) \sim -2\log|z-w|^2
\label{phiphi}\eeq
In the case of gravitational dressing, $\alpha_0{}^2<1$, while for the
$(p,q)$ models, $\alpha_0{}^2>1$ and $c_{p,q} = 1-6\frac{(p-q)^2}{pq}$.
In the case of a  gravitationally dressed $(p,q)$ model we have
$c_L+c_{p,q}=26$. 
In both models, the primary fields are vertex operators of the form 
\beq
V_\alpha (z,\bar{z})= :e^{i\alpha \phi(z,\bar{z})}:
\label{vertex}\eeq
with the conformal dimensions
\beq
h_\al = \al(\al-2\al_0).
\label{dims}\eeq
For the degenerate primary fields, $h$ and $\al$ takes the values:
\bea
h_{r,s}= \frac{(rq-sp)^2-(p-q)^2}{4pq} \nonumber \\
\al=\al_{r,s}=\frac{1}{2}(1-r)\al_+ + \frac{1}{2}(1-s)\al_-
\label{h_rs}\eea
where,
\beq
\al_\pm=\al_0 \pm \sqrt{\al_0^2+1}
\label{al_pm}\eeq
The operators (\ref{vertex}) have the usual OPE with the stress tensor:
\beq
T(z)V_\al(0)=\frac{h_\al}{z^2}V_\al(0) + 
\frac{1}{z}\partial V_\al(0) + \dots
\label{primary}\eeq
In minimal models, where $1\le r <q$ and $1\le s <p$, the exponential
primary 
fields (\ref{vertex}) and their descendants are the  only fields
in the theory. This changes if the primary field with the minimum
dimension, 
$h_{0,0}=h_{q,p}=-\frac{(p-q)^2}{4pq}$,
is included. This is particularly important for the $c_{p,1}$ models,
in which  the minimal model region ($1\le r <q$ and $1\le s <p$) is empty,
 and this field cannot be excluded. In these
cases the models are known to contain logarithmic operators as well as
ordinary primary fields \cite{f,gk}.

The exponential operator with the dimension $h_{0,0}$ has
$\al=\al_0$. For every other value of $\al$, there are two operators
with the same dimension  $h_\al$ given by eq. (\ref{dims}):
 $V_\al$ and $V_{(2\al_0-\al)}$. Since correlation functions can only
be non-zero if they satisfy the condition $\sum\al=2\al_0$, all correlation
functions have to  include one of the operators $V_{(2\al_0-\al)}$,
with all the $\al$'s given by eq. (\ref{h_rs}) \cite{dotsenko}. When
$\al=\al_0$, there are also two  primary operators with the 
same dimension; the second one is:
\beq
V_P = 
\left.\frac{\partial}{\partial \al} V_\al \right|_{\al=\al_0}
= i\phi(z)e^{i\al_0\phi(z)}.
\label{puncture}\eeq
This is called the puncture operator in the Liouville 
theory (in which $\al_0$ is imaginary).  
Since it contains $\phi$ and not just exponentials (or derivatives) of 
$\phi$, we might expect it to have logarithmic correlation functions,
but in fact 
$V_P$ is the only operator of this form which is an ordinary
primary operator, as can be seen by differentiating
eq. (\ref{primary}) with respect to $\al$, giving: 
\beq
T(z)\left\{\frac{\partial}{\partial \alpha}V_\al(0)\right\}=
\frac{\partial h_\al}{\partial \alpha}\frac{1}{z^2}V_\al(0) + 
\frac{h_\al}{z^2}\left\{\frac{\partial}{\partial \alpha}V_\al(0)\right\}
+\frac{1}{z}\partial 
\left\{\frac{\partial}{\partial \alpha}V_\al(0)\right\} + \cdots
\label{d/da}\eeq
This is the same as the OPE for a primary field only when 
\beq
\frac{\partial h_\al}{\partial \alpha}=2(\al-\al_0)=0,
\eeq
The ``puncture'' operator, with the minimum dimension,
is therefore an ordinary primary field, and has 
the usual $2$- and $3$-point
functions with no logarithms. However, it turns out that
there are logarithms in
$4$-point functions containing this operator. For example,
in the $c_{2,1}=-2$ model discussed in the previous section,
eqs. (\ref{charge}) and (\ref{h_rs}) give 
$\al_0=\al_{1,2}=1/\sqrt{8}$, so the ``puncture'' operator is 
just the $(1,2)$ operator. 
We can understand  why 
the logarithms appear in the four-point functions  by observing that   
 including the puncture operator in the model
naturally leads to the inclusion of other operators of the
form $i\phi e^{i\al\phi}$ (in this sense, the ``puncture'' operator
could be called a ``pre-logarithmic'' operator), as can be
seen by 
differentiating the OPE
\beq
e^{i\al\phi}(z,\zb)e^{i\be\phi}(0) \sim 
\frac{e^{i(\al+\be)\phi}(0)}{|z|^{2(h_\al+h_\be-h_{\al+\be})}}
\label{primeope}\eeq
giving:
\beq
i\phi e^{i\al_0\phi}(z,\zb)e^{i\be\phi}(0) \sim
\frac{i\phi e^{i(\al_0+\be)\phi}(0)}{|z|^{2(h_{\al_0}+h_\be-h_{\al+\be})}} +
2\be
\log|z|^2\frac{e^{i(\al_0+\be)\phi}(0)}
{|z|^{2(h_\al+h_\be-h_{\al+\be})}}
\label{ope+log}\eeq
Eq. (\ref{ope+log}) has the form of the OPE (\ref{OPE2})
that arises in the limit
where the dimensions of two operators become degenerate, leading to
four point functions with logarithmic singularities. There is then a second
operator, the logarithmic operator, for which instead 
of eq. (\ref{primary}) we find:
\beq
T(z)D_\al(0)=\frac{h_\al}{z^2}D_\al(0)
+\frac{1}{z^2}V_\al(0) +
\frac{1}{z}\partial D_\al(0) + \dots
\label{log-ope}\eeq
In the Coulomb gas picture, we can now see that the logarithmic
operator $D_\al$ can be written as:
\beq
D_\al = \left(\frac{\partial h_\al}{\partial \al}\right)^{-1}
\frac{\partial}{\partial \al}V_\al = 
\frac{i}{\al-\al_0}\phi e^{i\al\phi}
\eeq
Note that this operator cannot be defined when $\al=\al_0$. 

It is interesting to examine when the puncture operator, and therefore
also logarithmic operators, will appear
in a gravitationally dressed conformal field theory. In this case, the
gravity sector is described by the action (\ref{action}) with $\al_0$
chosen to give the total central charge $26$ \cite{gn,ddk},
 so that if a model with
central charge $c_{p,q}$ is coupled to gravity, we have
\beq
\al_0^2=-1-\frac{(p-q)^2}{4pq}
\eeq
Since $\al_0$ is imaginary, the vertex operators of the Liouville
model
are $V_\beta=e^{\beta\phi}$, with $\beta$ real, and with dimensions
\cite{gn}
\beq
h_\beta=-\beta(\beta-2i\al_0)
\eeq
The puncture operator, $V_P=\phi e^{i\al_0\phi}$, therefore
has in this case the maximum rather than the minimum 
dimension 
\beq
h_P = h_{i\al_0}=1 + \frac{(p-q)^2}{4pq}
\label{h_P=}\eeq
Primary fields $\Phi_{r,s}$
from the $(p,q)$ model, with dimension $h_{r,s}$ are
dressed by fields from the Liouville model with dimension $h_\beta$,
to form composite fields with $\Phi_{r,s}e^{\beta\phi}$ with
total conformal dimension $1$, so that it makes sense to integrate
the dressed fields over the surface \cite{ddk}. We therefore have:
\beq
h_\beta + h_{r,s} =1
\label{h+h=1}\eeq
In a $c=1$ model
coupled to gravity, the puncture operator is  a cosmological 
constant operator $(h_P=1)$, and  logarithmic operators 
do exist in that model
\cite{bilal}. 
From the above discussion, 
we expect that  there will also be logarithmic operators 
in any gravitationally dressed $c<1$ model if the puncture operator
appears as the dressing of one of the primary fields in the matter
theory.
From eqs. (\ref{h+h=1}) and (\ref{h_P=}), we can see that the field 
that will be dressed by the puncture operator has the dimension
\beq
1-h_P=-\frac{(p-q)^2}{4pq}=h_{0,0}
\eeq
The puncture operator therefore appears as the dressing of   the
``puncture'' (or pre-logarithmic)
operator in the $(p,q)$ model. We therefore expect that there will be
no logarithmic operators in minimal models coupled to gravity, because
there are no ``puncture'' (ie. no pre-logarithmic) operators, but
that
when logarithmic theories are
coupled to gravity, there will be additional
 logarithmic operators  in
the gravity sector. The only case in which
logarithmic operators exist in the gravitationally dressed
model but not in the model without gravity is $c=1$. This is because
in a $c=1$ model (without gravity), the field with the minimum
dimension is the identity, but there is no ``puncture'' operator
with the same dimension.

\section{Correlation Functions}
We now consider correlation functions in models which contain both
operators $V_\al$ and $D_\al$.   As usual,
the $2$-point functions are completely determined by projective
invariance, and they can  be derived either from the four-point
functions
as in \cite{gur,tsvelik}, or from the transformation law implied by
 eq. (\ref{log-ope}), as in \cite{rtak} (correlation functions can
also be found using the ${\cal W}_\infty$ algebra \cite{shaf}): 
\bea
z\rightarrow z+\epsilon(z)&& \nonumber \\
\delta D_\al(z) &=& \partial \epsilon (z)
\left[h_\al D_\al (z) + V_\al (z) \right] +
\epsilon (z) \partial D_\al (z) \nonumber \\
\delta V_\al(z) &=& \partial \epsilon (z)
h_\al V_\al (z)  +
\epsilon (z) \partial V_\al (z) 
\eea
Correlation functions must be invariant under the projective 
transformations, which can be written as \cite{rtak}
\bea
\left[ L_n,V_\al(z) \right] &=& z^{n+1}\partial V_\al 
+(n+1)z^nh_\al V_\al \nonumber \\
\left[ L_n,D_\al(z) \right] &=& z^{n+1}\partial D_\al 
+(n+1)z^nh_\al D_\al +(n+1)z^nV_\al\nonumber \\
&&n=0,\pm1
\label{proj}\eea
and non-zero correlation functions
must also satisfy the neutrality condition $\sum_i\al_i=2\al_0$.
There are similar relations for the $\bar z$ dependence.
For the ordinary primary fields $V_\al$, we find as usual:
\beq
\langle V_\al(z,\zb)V_{2\al_0-\al}(0)\rangle = \frac{A}{|z|^{4h_\al}}
\label{nonzero2pt}\eeq
In ordinary conformal field theories, the constant $A$ is not
determined,
but if  the logarithmic operator $D_\al$ exists, eq. (\ref{proj})
leads to the following equations for 
$\langle V_\al(z)D_{2\al_0-\al}(w)\rangle$: 
\bea
&\left[ \partial_z+\partial_w \right]
\langle V_\al(z)D_{2\al_0-\al}(w)\rangle=0 \nonumber\\
&\left[z\partial_z+w\partial_w+2h_\al\right]
\langle V_\al(z)D_{2\al_0-\al}(w)\rangle
+\langle V_\al(z)V_{2\al_0-\al}(w)\rangle=0 \nonumber\\
&\left[z^2\partial_z+w^2\partial_w+2h_\al(z+w)\right]
\langle V_\al(z)D_{2\al_0-\al}(w)\rangle
+2w\langle V_\al(z)V_{2\al_0-\al}(w)\rangle=0 
\label{2pteqs}\eea
These equations are only consistent if $A=0$, so when the 
logarithmic operator $D_\al$ exists,we must have:
\beq
\langle V_\al(z,\zb)V_{2\al_0-\al}(0)\rangle = 0
\label{zero2pt}\eeq
Solving eq. (\ref{2pteqs}), and the similar equations for
$\langle D_\al(z,\zb)D_{2\al_0-\al}(0)\rangle$ then leads to:
\bea
\langle V_\al(z,\zb)D_{2\al_0-\al}(0)\rangle &=& 
\frac{B}{|z|^{4h_\al}}
\nonumber \\
\langle D_\al(z,\zb)D_{2\al_0-\al}(0)\rangle &=& 
\frac{-2B\log|z|^2 +\delta}{|z|^{2h_\al}}.
\label{log2pt}\eea
 Since $V_P(z)$ is an ordinary primary field, and does
not have a logarithmic partner, we have:
\beq
\langle V_{\al_0}(z,\zb)V_P(0) \rangle =
\frac{1}{|z|^{2h_{\al_0}}}
\label{prime2pt}\eeq
However, we can also compute all these two-point functions
directly using eqs. (\ref{phiphi}) and (\ref{vertex}), giving
\bea
\langle V_\al(z,\zb)V_{2\al_0-\al}(0)\rangle &\sim& 
\frac{\langle V_{2\al_0}\rangle}{|z|^{2h_\al}} \nonumber \\
\langle V_{\al}(z,\zb)D_{2\al_0-\al}(0) \rangle 
\sim \langle V_{\al_0}(z,\zb)V_P(0) \rangle
&\sim&
\frac{\langle D_{2\al_0}\rangle}{|z|^{2h_{\al_0}}}
\label{wrong2pt}\eea
Since $V_{2\al_0}$ is an identity operator ($h_{2\al_0}=0$), in
ordinary CFT we would take $\langle V_{2\al_0}\rangle=1$, but if any
of the operators $D_\al$ or $V_P$ exist, eqs. (\ref{zero2pt}), 
(\ref{log2pt}),
(\ref{prime2pt}) and (\ref{wrong2pt}) will only be consistent if we
have:
\bea
\langle V_{2\al_0} \rangle = \langle I \rangle &=& 0 \nonumber \\
\langle D_{2\al_0} \rangle &=& 1 \label{right2pt}\eea
It can then be checked that, assuming eq. (\ref{right2pt}),
we find using eq. (\ref{phiphi}) that
\beq
\langle D_{2\al_0}(z,\zb)  D_0(0)
\rangle = -2\log|z|^2
\eeq
In agreement with eq. (\ref{log2pt}).

There must therefore be at least one logarithmic operator, with
dimension $0$, in any model which contains the puncture operator,
 which is indeed the case in th $c=-2$ and other $c_{p,q}$
models \cite{gur,f,k,gk}. In general there will also be 
logarithmic operators with
other dimensions, but it is necessary to actually compute the
four-point functions or fusion rules to determine for which values
of $\al$ the operators $D_\al$ exist. The four-point functions
containing the puncture operator are given by the expressions given
in \cite{dotsenko}, but as these integrals diverge they have to be
analytically continued from values of $c$ for which
$\al_{q,p}\neq\al_0$, and it is in this way that the logarithmic
singularities can appear. These correlation functions always contain
exactly one operator $V_{2\al_0-\al_{q,p}}$, in addition  to ordinary
operators $V_{\al_{r,s}}$ and screening operators, in order to satisfy
the condition $\sum \al =2\al_0$. In the limit where
$\al_{q,p}\goto \al_0$, we could take either 
\beq
V_{2\al_0-\al_{q,p}}\goto
V_{\al_0}
\label{limit1}\eeq
 or 
\beq
V_{2\al_0-\al_{q,p}}\goto V_P.
\label{limit2}\eeq
 The logarithms in the
four- point functions tell us that eq. (\ref{limit2}) is correct. In
addition, any correlation functions containing two or more puncture
operators can be seen to be the analytic continuation of functions
with $\sum \al \neq 2\al_0$, and must therefore vanish. This is
important as calculating these functions using eq. (\ref{phiphi})
leads to extra factors of $\log |z|$ which should not appear.

We might expect from eqs. (\ref{primeope}) and  (\ref{ope+log}) that
we would obtain  a logarithmic operator from the fusion of the
puncture operator with any other primary operator. However, this is
not always 
the case in the $c_{p,1}$ models.  We can see why this is by
observing that in some cases the OPE of the two primary fields only
contains one primary field, as can be seen from eq. (\ref{minope}):
\beq
\Phi_{1,p}(z)  \Phi_{s,1}(0) \sim z^{h_{s,p}-h_{1,p}-h_{s,1}}
 \Phi_{s,p}(0) +\mbox{descendants}
\label{1pxs1}\eeq
This means that the relevant four point function has only one
conformal block. The logarithmic operators appear when two of the
dimensions of fields in the OPE become degenerate -- there is no
way for this to happen
 in this case. However, as we have seen, the OPE of
$\Phi_{1,p}$ with itself does contain the logarithmic operator $D_0$:
\beq
\Phi_{1,p}(z) \Phi_{1,p}(0) \sim  z^{-2h_{1,p}}\left[
D_0(0) + \log|z|^2 I +\cdots \right] +\dots
\label{1px1p}\eeq
This implies that logarithms must also appear in correlation functions
of primary fields with dimension $h_{s,p}$, even if they do not
contain the puncture operator, as can be seen by writing:
\bea
\Phi_{s,p} \times \Phi_{s,p} &\sim& \Phi_{s,1} \times \left[\Phi_{1,p}
\times \Phi_{1,p}\right] \times \Phi_{s,1} \nonumber \\ &\sim&
\Phi_{s,1}  \times \left[D_0(0) + \log|z|^2 I \right] 
\times \Phi_{s,1}
\label{spxsp}\eea
Another way to see this is to observe that the OPE 
$\Phi_{s,p}(z)\Phi_{s,p}(0)$ contains two primary fields, the identity
and $\Phi_{1,2p-1}$ which have the same dimension $h_{1,2p-1}=0$
when $c=c_{p,1}$. The indicial equation for the corresponding
differential equation will therefore have two equal roots, and so 
there will be logarithms in one
of the conformal blocks. Therefore all the $\Phi_{s,p}$ operators,
not just the $\Phi_{1,p}$ operator will behave as ``pre-logarithmic''
operators in the $(p,1)$ model.

\section{Logarithmic Operators Degenerate with Descendants}

So far we have considered the logarithmic operators which occur when
the dimensions of two primary fields become degenerate. for example,
in the $c=-2$ model, the dimensions $h_{1,1}=h_{1,3}=0$, giving the
logarithmic pair with dimension $0$. This is the situation when the
roots of the indicial equation for the conformal blocks are equal.
Logarithmic operators can also
occur when two of the dimensions in the spectrum differ by an
integer, so that one primary field becomes
degenerate with a descendent of another,
and this type of logarithmic operator
 also appear in $c_{p,q}$ models. This occurs when the roots of
 the indicial
equation differ by an integer.
 In this case, we have a primary operator $C^1$ with dimensions
$(h-n,h)$, ($n$ is an integer), and with a null vector on the level $n$,
and another primary operator $\bar{C}^1$ with dimensions $(h,h-n)$ 
(up to now, all
the operators we have considered have had equal left and right
dimensions).
There is also a primary operator $C$ and a logarithmic operator $D$ with
dimensions $(h,h)$, with the relations \cite{gk}:
\bea
\sigma_{-n} C^1 &=& C \nonumber \\
\bar{\sigma}_{-n} \bar{C}^1 &=& C \nonumber \\
L_0 D &=& h D + C \nonumber \\
(L_1)^n D &=& \beta C^1
\label{morelogs}\eea
Where $\sigma_{-n} $ is the combination of Virasoro generators $L_i$
which gives the null vector and $\beta$ is a constant.
In addition, $C$ and $C^1$ satisfy the usual relations for primary
operators. We can also redefine $D(z)$ by adding descendants of
$C^1(z)$ so that
\beq
L_iD = 0, ~~~~~~~~~~~~~~~i\geq 2.
\eeq
This leads to the following OPE:
\beq
T(z)D(0) \sim  \frac{L_1D(0)}{z^3}
+\frac{h D(0)}{z^2} +\frac{C(0)}{z^2} +\frac{\partial D(0)}{z} +\cdots
\label{morelogope}\eeq
As before, the operators with this behaviour  have the form of
derivatives with respect to $\al$ of ordinary operators.
 Also, since  $C^1$  must necessarily have $h \neq
\bar h$, we write $\phi$ as:
\beq
\phi(z,\zb) = \varphi(z) + \bar{\varphi}(\zb)
\eeq 
The simplest example occurs for $h=1$. In this case we have three
primary fields; $C$ with conformal weights $(1,1)$, $C^1$ with weights
$(0,1)$ and $\bar C^1$ with weights $(1,0)$; and a logarithmic
operator $D$ with weights $(1,1)$, which are constructed as follows:
\bea
C^1 &=& :e^{i\al_\pm \bar{\varphi}(\zb)}: \nonumber \\
\bar C^1 &=& :e^{i\al_\pm \varphi(z)}: \nonumber \\ 
C &=& :e^{i\al_\pm \phi(z,\zb)}: \nonumber \\
D &=& \frac{d}{d\al}\left\{ e^{i(\al_\pm+\al) \phi(z,\zb))}
+\lambda L_{-1}e^{i[\al_\pm \bar{\varphi}(\zb) +\al\phi(z,\zb)]}
+\lambda \bar{L}_{-1}e^{i[\al_\pm \varphi(z) +\al\phi(z,\zb)]}
\right\}_{\al=0} \nonumber \\
&=&i:\left[\phi(z,\zb) e^{i\al_\pm \phi(z,\zb))}
+\lambda \partial \phi(z,\zb)e^{i\al_\pm \bar{\varphi}(\zb)}
+\lambda \bar{\partial}\phi(z,\zb)e^{i\al_\pm \varphi(z)}\right]:
\eea
where $\lambda$ has to be chosen to give the correct value of $\beta$
in eq. (\ref{morelogs}). More generally, when the dimensions differ by
an integer $n$, the logarithmic operator with the behaviour of 
eq. (\ref{morelogope}) can be written as:
\beq
D = \frac{d}{d\al}\left\{:Ce^{i\al\phi}: 
+\lambda\sigma_{-n}:C^1e^{i\al\phi}:\right\}_{\al=0}
 +\mbox{descendants of }C^1 +\mbox{c.c.}
\eeq

\section{WZNW Models for $SU(2)_0$ and $SL(2)$}

Among the other models in which there are logarithmic operators are
the WZNW model on the group $SU(2)$ at $k=0$, and possibly also the
$SL(2)/U(1)$ coset model. To study these models in the same way as the
$c_{p,q}$ models, we use the free field representation of
\cite{gerasimov}.
We introduce three free fields, $u$, $v$ and $\phi$, each of which
obeys eq. (\ref{phiphi}). The stress tensor for the WZNW model at
level $k$ is
written as:
\beq
T(z)=T_u(z)+T_v(z)+T_\phi(z)
\label{WZWT} \eeq
where:
\bea
T_\varphi=
-\frac{1}{4}:\partial_z\varphi\partial_z\varphi:+
i\al_{0,\varphi}\partial_z^2\varphi \nonumber \\
\varphi=u,~v,~\phi \nonumber \\
\al_{0,u}^2=-\frac18,~~~~~\al_{0,u}^2=\frac18,
~~~~~\al_{0,\phi}^2=\frac{1}{4(k+2)}
\eea
The currents, $J^\pm$ and $J^0$, are:
\bea
J^+ &=& \frac1{\sqrt2} \partial v e^{(-u+iv)/\sqrt{2}} \nonumber \\
J^0 &=& -\frac{i\sqrt{(k+2)}}{2}\partial\phi + \frac{1}{\sqrt{2}} 
\partial u
\nonumber \\
J^- &=& \frac1{\sqrt2}\left[ -\sqrt{2(k+2)}\partial\phi
-i(k+2)\partial u - (k+1)\partial v \right]e^{(u-iv)/\sqrt{2}}
\label{currents} \eea
The primary fields of  the WZNW model, with dimensions
$h_j=\frac{j(j+1)}{k+2}$, are the vertex operators:
\beq
V_{j,m}= e^{-ij\phi/\sqrt{k+2}}e^{\sigma(u-iv)},
~~~~~~~\sigma=\frac{m-j}{\sqrt{2}}
\label{WZWvertex}\eeq
The OPEs of these operators with the currents are:
\bea
J^+(z)V_{j,m}(0)&=& \frac{i(m-j)}{z} V_{j,m+1}(0) \nonumber \\
J^0(z)V_{j,m}(0)&=& \frac{m}{z} V_{j,m}(0) \nonumber \\
J^-(z)V_{j,m}(0)&=& \frac{i(j+m)}{z} V_{j,m-1}(0)
\label{JV}\eea

\subsection{Jordan Blocks in Affine Algebra for $SL(2)$}
As in the $c_{p,q}$ models, non-zero correlation functions in the 
free field formulation of the WZNW model must 
contain one of the operators $V_{-1-j,m}$, which has the same
conformal dimension as $V_{j,m}$. The equivalent of the puncture
operator therefore has $j=-1-j$, so $j=-1/2$. of course this operator,
which is in an infinite dimensional representation of $SU(2)$,
cannot exist in the WZNW model on the group $SU(2)$, but it could
exist in the model based on the non-compact group $SL(2,{\cal
R})$. This is obtained by simply redefining $J^\pm$ as $iJ^\pm$ in
eq. (\ref{currents}).   By analogy with the $c_{p,1}$ models, we might
therefore expect the WZNW model with the non-compact group to include
operators of the form
\beq
\tilde{V}_{j,m} = \frac{\partial}{\partial j} V_{j,m}
= -\frac{i\phi}{\sqrt{k+2}}
e^{-ij\phi/\sqrt{k+2}}e^{\sigma(u-iv)}
\eeq
We can find the OPEs of these new operators
 with 
the currents either by differentiating eq. (\ref{JV}) or
using eqs. (\ref{currents}) and (\ref{phiphi}).
The result is (taking into account the extra factor of $i$ in 
$J^\pm$):
\bea
J^+(z)\tilde{V}_{j,m}(0)&=& \frac{(m-j)}{z} \tilde{V}_{j,m+1}(0) 
\nonumber \\
J^0(z)\tilde{V}_{j,m}(0)&=& \frac{m}{z} \tilde{V}_{j,m}(0) 
+\frac{1}{z} V_{j,m}(0) \nonumber \\
J^-(z)\tilde{V}_{j,m}(0)&=& \frac{(j+m)}{z} \tilde{V}_{j,m-1}(0)
+\frac{2}{z} V_{j,m-1}(0)
\label{JtildeV}\eea

From eqs. (\ref{JV}) and (\ref{JtildeV}), we can see that, just as 
the logarithmic operators $D_\al$ together with the ordinary
primary operators formed Jordan blocks for $L_0$, the $V$ and 
$\tilde{V}$
operators form Jordan blocks for the zero-modes of the 
currents:
\bea
J^0_0V_{j,m}=mV_{j,m} &~~~~~~~~~~~~~~~~&
J^0_0\tilde{V}_{j,m}=m\tilde{V}_{j,m} +V_{j,m}\nonumber \\
J^+_0V_{j,m}=(m-j)V_{j,m+1} &~~~~~~~~~~~~~~~~&
J^+_0\tilde{V}_{j,m}=(m-j)\tilde{V}_{j,m+1}
\nonumber \\
J^-_0V_{j,m}=(m+j)V_{j,m-1} &~~~~~~~~~~~~~~~~&
J^-_0\tilde{V}_{j,m}=(m+j)\tilde{V}_{j,m-1}+2V_{j,m-1}  
\eea
However, it turns out that the $\tilde{V}$'s are not 
logarithmic operators, at least in the WZNW model, as we can see by 
calculating the OPE with the stress tensor, which has the Sugarawa
form:
\beq
T(z)=\frac{1}{k+2} :J^a(z)J^a(z): = \frac{1}{k+2}:\left(
J^0J^0-\frac12J^+J^--\frac12J^-J^+ \right):
\eeq
When we calculate $T(z)\tilde{V}_{j,m}(0)$ using eq. (\ref{JtildeV}),
the $V_{j,m}$ terms all cancel, leaving the OPE for a primary field of
the Virasoro algebra:
\beq
T(z) \tilde{V}_{j,m}(0) = \frac{j(j+1)}{k+2}
\frac{\tilde{V}_{j,m}(0)}{z^2} + O\left(\frac{1}{z}\right)
\eeq
In fact, we can see that this had to be true,  because 
$L_0=J^a_0J^a_0/(k+2)$ is just the Casimir operator for
$SL(2)$ and must therefore be diagonalizable, so there can be no 
non-trivial Jordan blocks.

However, the $\tilde{V}$'s may become logarithmic operators in models
with modified stress tensors, in which $L_0$ is not
a Casimir operator. There are two examples of such models in
which logarithmic operators do exist \cite{oleg}. One is
2 dimensional gravity, for which the stress tensor can be written as
\cite{kpz}:
\beq
T(z) =  \frac{1}{k+2}:J^a(z)J^a(z): + \frac{\partial}{\partial z}
J^0(z)
\eeq
If we compute the OPE of $\tilde{V}_{j,m}$ with this stress tensor,
the first term, which is just the Sugarawa form, again only gives
$\tilde{V}_{j,m}$, but the second term gives us the mixing
with $V_{j,m}$ which characterizes logarithmic operators:
\beq
T(z) \tilde{V}_{j,m}(0) = \left(\frac{j(j+1)}{k+2}-m\right)
\frac{\tilde{V}_{j,m}(0)}{z^2} - m\frac{V_{j,m}(0)}{z^2}
+ O\left(\frac{1}{z}\right)
\eeq
This has the form of the OPE for a logarithmic operator $D$ (eq. 
(\ref{TD})), with $D=\tilde{V}_{j,m}$ and $C=-mV_{j,m}$. The
logarithmic operators found in \cite{bilal} are examples of this type
of operator.

The second model in which this happens
 is the $SL(2)/U(1)$ coset model, which describes the Witten
2D black hole \cite{Witten}, with the stress tensor:
\beq
T(z) =  \frac{1}{k+2}:J^a(z)J^a(z): - \frac{1}{k}:J^0(z)J^0(z):
\eeq
As before, the second term leads to mixing between $\tilde{V}_{j,m}$
and $V_{j,m}$:
\beq
T(z) \tilde{V}_{j,m}(0) = \left(\frac{j(j+1)}{k+2}
-\frac{m^2}{k}\right)
\frac{\tilde{V}_{j,m}(0)}{z^2} - 
\frac{2m}{k}\frac{V_{j,m}(0)}{z^2}
+ O\left(\frac{1}{z}\right)
\eeq
Again, this has the form of eq. (\ref{TD}) with $D=k\tilde{V}_{j,m}$
 and $C=-2mV_{j,m}$. It is therefore possible for logarithmic 
operators to exist in a coset model even if they did not exist 
in the original WZNW model.

\subsection{Logarithmic Operators in the 
WZNW Model for $SU(2)$ at $k=0$}
From the above discussion, we can see that the 
only way it is possible for logarithmic operators to appear in a
WZNW model is if they are not annihilated by all positive modes of the
currents. For example, if
\beq
J^a_1 D \neq 0
\eeq
then $L_0$ will not be a Casimir operator:
\beq
L_0 = \frac{1}{k+2} :J^a_n J^a_{-n}: = \frac{1}{k+2}\left[
J^a_0J^a_0 +2 J^a_{-1} J^a_1 \right]
\eeq
We will therefore have a representation
of the  algebra of the type (\ref{KMjordan2})
instead of (\ref{KMjordan1}).
These logarithmic operators will  be of the type discussed in
the previous section with
\beq
L_1 D = \frac{2}{k+2} J^a_0J^a_1 D \neq 0
\eeq
One model in which this happens is the WZNW model for $SU(2)$ at $k=0$.
To see why logarithmic operators should be expected in this model, we
concentrate on the $\phi$ dependent parts of the stress tensor $T(z)$
and primary operators $V_{j,m}$, since the $u$ and $v$ dependent parts
are independent of $k$. The $\phi$ dependent part
 of the stress tensor $T(z)$ is in fact identical to the stress tensor
for the $c=-2$ model, as can be seen by comparing eqs. (\ref{charge})
and (\ref{T(z)}) with:
\beq
T_\phi(z)= -\frac14:\partial_z \phi  \partial_z \phi: - 
\frac{i}{\sqrt 8}\partial_z^2 \phi .
\eeq
 The primary fields in the two models (without the $u$ and $v$
dependent parts) are also the same: comparing eqs. (\ref{h_rs})
and (\ref{WZWvertex}), we see that
\bea
V_j(k=0) &\sim& V_{j+1,1}(c=-2) ~~~~~~j=0,1,\dots \nonumber \\
V_j(k=0) &\sim& V_{j+\frac32,2}(c=-2) ~~~~~~j=\frac12,\frac32,\dots 
\eea
where $V_{r,s}(c=-2)$ is the operator with dimension $h_{r,s}$ in the
$c=-2$ model, and the dimensions $h_j$ and $h_{r,s}$ are equal.
 The puncture operator with dimension $h_{1,2}=-\frac18$ 
therefore corresponds to $j=-\frac12$ and
does not exist in the WZNW model for $SU(2)$, but, as
discussed earlier, logarithmic operators also appear in the OPE of any
of the fields with dimension $h_{r,2}$ in the $c=-2$ model, and so we
expect them to appear also in the OPE of any of the fields with
half-integer $j$ in the $k=0$ model. The case of $j=1/2$ was studied
in \cite{cklt}, where the following OPE was found:
\bea
g_{\epsilon_1 \bar{\epsilon}_1} (z_1, \bar{z}_1) g^{\dagger}_{\bar{\epsilon_2}
\epsilon_2} (z_2, \bar{z}_2) = |z_{12}|^{-3/2}
&\times& \left\{ z_{12} 
\delta_{\bar{\epsilon}_1 \bar{\epsilon}_2} t^i_{\epsilon_1
\epsilon_2} K^i (z_2) +  \bar{z}_{12} \delta_{\epsilon_1
\epsilon_2} \bar{t}^i_{\bar{\epsilon}_1 \bar{\epsilon}_2}
\bar{K}^i (\bar{z}_2)    \right. \label{gg}\\
&&+ \left. |z_{12}|^2  t^i_{\epsilon_1
\epsilon_2} \bar{t}^j_{\bar{\epsilon}_1 \bar{\epsilon}_2}
\left[ D^{ij} (z_2, \bar{z}_2) + \ln |z_{12}| C^{ij} (z_2, \bar{z}_2)
\right] + \cdots \right\} \nonumber
\eea
Here $K$ and $\bar K$ are primary fields with dimensions $(1,0)$ 
and $(0,1)$, and $C$ and $D$ are the logarithmic pair with dimensions 
$(1,1)$. From the above discussion we expect to find $J^a_1D\neq 0$,
and we can check this using the OPE for the  currents with the
primary field $g$:
\beq
J^a(w) g_{\epsilon_1 \bar{\epsilon}_1} (z_1) 
g^{\dagger}_{\bar{\epsilon_2}\epsilon_2} (z_2) =
\frac{1}{w-z_1}t^a_{\epsilon_1 \eta_1}g_{\eta_1 \bar{\epsilon}_1} (z_1) 
g^{\dagger}_{\bar{\epsilon_2}\epsilon_2} (z_2)
+\frac{1}{w-z_2}g_{\epsilon_1 \bar{\epsilon}_1} (z_1) 
g^{\dagger}_{\bar{\epsilon_2}\eta_2} (z_2)t^a_{\eta_2 \epsilon_2} +\cdots
\label{jgg}\eeq
Taking the limit as $z_1\goto z_2$ in eq. (\ref{jgg}) using 
eq. (\ref{gg}), we find:
\beq
J^a(w)t^i_{\epsilon_1\epsilon_2}D^{ij} (z,\bar{z}) = 
\frac{t^a_{\epsilon_1 \eta_1}t^i_{\eta_1\epsilon_2}}{(w-z)^2}
\bar{K}^j (\bar{z})
+ \frac{f^{aib}}{w-z}t^b_{\epsilon_1\epsilon_2}D^{bj} (z,\bar{z})
+ \cdots
\eeq
from which we can see that, as expected, $J^a_1D\sim\bar{K}$.

As was pointed out in \cite{cklt}, $K$ and $\bar{K}$ are conserved
currents, indicating that the WZNW model at $k=0$ has an additional
symmetry (as well as the $SU(2)$ symmetry), which we 
can also try to
understand in terms of the relation with the $c=-2$ model.
 The extended symmetry in the $c=-2$ model
has a ${\cal W}$ algebra, which is generated by the dimension $3$
fields, $\Phi_{3,1}$ \cite{kausch}.  
There are dimension $3$ fields in the $k=0$
model as well (the primary fields with $j=2$), 
and we might conjecture that these also generate a ${\cal W}$ algebra.
However, the operator product expansions, and therefore the algebra,
of the $j=2$ fields in the $k=0$ model and the $\Phi_{3,1}$ fields in 
the $c=-2$ model are not the same.  In the $c=-2$ model, we
have:
\beq
\Phi_{3,1} \times \Phi_{3,1} \sim \left[\Phi_{1,1}\right]
+ \left[\Phi_{3,1}\right] + \left[\Phi_{5,1}\right]
\label{31x31}\eeq
while in the WZNW model, we have: 
\beq V_2 \times V_2 \sim \left[V_0\right] + 
\left[V_1\right] + \left[V_2\right] + \left[V_3\right] + 
\left[V_4\right] 
\label{2x2}\eeq
The appearance of $V_1$ in eq. (\ref{2x2}), when the corresponding
operator $\Phi_{2,1}$ does not appear in eq. (\ref{31x31}), will
lead to extra terms in the singular part of the OPE for
$V_2(z)V_2(0)$,
so that the algebra of these operators is not the same as the algebra of
the
$\Phi_{3,1}$ operators in the $c=-2$ model.
The reason why the correspondence between operators in the two models 
does not extend to the OPEs (or correlation functions) is
that the screening operators $Q=\oint J$
used to construct correlation functions are different. 
In the $c=-2$ model, 
$J=e^{i\al_\pm \phi}$,
where $\al_\pm$ is given by eq. (\ref{al_pm}), while  in the $k=0$
model
we have a completely different expression
$J= -ie^{(-i\phi-u+iv)/\sqrt{2}}\partial v$.
It remains a difficult problem to determine the full symmetry algebra
of the $k=0$ model.

\section{Conclusion}

In this paper, we have shown that the logarithmic operators which 
were known to exist in several different conformal field theories
-- $c_{p,1}$ models, gravitationally dressed CFT and some WZNW and
coset models -- can be understood in the free field formulation
of the models as originating from a ``pre-logarithmic'' operator, 
the puncture operator. The logarithmic operators are
derivatives
of the ordinary primary vertex operators,  while the puncture
operator  is the only  ordinary primary operator that has the
form of a logarithmic operator. The existence of the
puncture operator leads naturally to the existence of
logarithmic operators, and this gives us an easy way to guess
when
logarithmic operators will appear in other models, where it may be
difficult
or impossible to explicitly compute the conformal blocks.

We have also shown that in some models there may be operators which
form Jordan blocks for the Kac-Moody algebra, in the same way as the
logarithmic operators form Jordan blocks for the Virasoro algebra.
These occur in physically interesting models, such as gravitationally
dressed CFT and the model of 2D black holes. The further investigation
of this subject is of great interest.

\end{document}